\documentclass[prl,aps,preprint,superscriptaddress,nofootinbib]{revtex4}
\usepackage{graphicx} % Include figure files
\usepackage{epsfig}
\usepackage{dcolumn}  % Align table columns on decimal point
\usepackage{bm}       % bold math

\def\ee{{\rm e}}  \def\ii{{\rm i}}  \def\xx{\hat{\bf x}}  \def\yy{\hat{\bf y}}  \def\zz{\hat{\bf z}}
\def\rb{{\bf r}}  \def\qb{{\bf q}}  \def\ub{{\bf u}}
\def\Eb{{\bf E}}  \def\Fb{{\bf F}}  \def\pb{{\bf p}}

\begin{document}

\title{Collective oscillations in optical matter}

\author{F. J. Garc\'{\i}a de Abajo}
\address{Instituto de \'Optica - CSIC, Serrano 121, 28006 Madrid, Spain \\ E-mail: jga@io.cfmac.csic.es}

\begin{abstract}
Atom and nanoparticle arrays trapped in optical lattices are shown
to be capable of sustaining collective oscillations of frequency
proportional to the strength of the external light field. The
spectrum of these oscillations determines the mechanical stability
of the arrays. This phenomenon is studied for dimers, strings, and
two-dimensional planar arrays. Laterally confined particles free to
move along an optical channel are also considered as an example of
collective motion in partially-confined systems. The fundamental
concepts of dynamical response in optical matter introduced here
constitute the basis for potential applications to quantum
information technology and signal processing. Experimental
realizations of these systems are proposed.
\end{abstract}
\maketitle

% --- text ----------------------------------------------------

\section{Introduction}

Periodic standing light patterns created by coherent laser beams
offer the possibility to trap objects ranging from single atoms
\cite{GME02} to micron-size particles \cite{BFG1989,BFG1990},
forming ordered arrays that can be viewed as artificial crystals
with the periodicity of the optical field. Starting from the
equilibrium configuration of an ordered array with one particle per
lattice site, the perturbation produced by slightly displacing one
of the particles will be transmitted to the others via their mutual
electromagnetic interaction, which is a complex function of the
particles positions. This results in collective motion of the
particles that resembles phonon vibrations in crystals and that we
regard as collective oscillations in optical matter.

In this context, the interaction of ultracold atoms with optical
lattices has recently attracted considerable attention because of
its potential use for quantum information technology, in which the
realization of trapping of no more than one atom per lattice site
has been an important benchmark \cite{GME02}. Using light tuned
close to some atomic resonance, strong light-induced interaction
between atoms trapped in different sites can take place even for
commonly employed lattice periods of the order of 1 $\mu$m, as
suggested by recent observations in optical cavities under
atom-cavity resonance conditions \cite{MFM00,MEH03}. These
interatomic interactions are the driving force of the collective
oscillations considered here, which for reasonably-cold atoms offer
a potentially practical realization of quantum gates \cite{JCZ00}
involving large numbers of qubits.

Early attempts to bind small particles using light forces
\cite{A1970,AD1975,A1980,AD1987,BFG1989,BFG1990} led to the
development of optical tweezers \cite{A1970,AD1987}, capable of
trapping and aligning objects ranging from micro-organisms
\cite{AD1987} to metallic nanoparticles \cite{HBH05,PLK06}.
Furthermore, holographic tweezers have been developed to trap
tailored arrays of micron-size objects \cite{G03,GR06} that offer an
excellent playground to test many of the concepts discussed below.
Manipulation of micro-particles using plasmons has been recently
demonstrated as well \cite{RZG07}, whereas fine tuning of
nanoparticle positions has been theoretically proved to be
realizable by coupling to plasmonic nanostructures \cite{paper124}.

Mutual inter-particle interaction induced by external illumination
has been shown to lead to interesting effects such as
configurational bistability in pairs of spheres \cite{MDW06} and
crystallization in linear particle arrays trapped at the focus of
counter-propagating lasers \cite{TCD02}. A close relative of the
oscillations studied here are those of colloidal crystals
\cite{HV98}, mediated by electrostatic interaction similar to
phonons in ionic crystals.

% Control over individual particles has been achieved using optical tweezers \cite{NBX97}, and several alternative trapping mechanisms have been proposed \cite{OK99,CRN02}. The rotational motion of the particles can be controlled as well \cite{GO01,FNH98}.
% Micro-rotors driven by light: when the light carries angular momentum (elliptical polarization) \cite{GO01} or for linear polarization and helical shapes \cite{FNH98}.
% Atoms moving in light standing waves have been shown to describe Bloch states similar to electrons evolving in periodic crystal potentials \cite{BPR96}. Under the appropriate low-temperature conditions, Bose-Einstein condensates can be formed all-optically in such lattices \cite{BSC01}.
% Recent calculations: optical tweezers\cite{NBX97}.
% Recent calculations: trapping near a small hole \cite{OK99}.
% Recent calculations: manipulation with a probe \cite{CRN02}.
% biding between two particles and anisotropic forces in a nematic colloid \cite{YYY04}
% binding and alignment of gold nanorods \cite{PLK06}
% trapping of gold nanoparticles \cite{HBH05}
% photonic crystal band gaps based upon atoms \cite{VST96}

Here, we examine the collective oscillation modes of atoms and
particles trapped in optical lattices. For atomic lattices, strong
inter-atomic interaction effects are predicted in the oscillation
spectrum near atomic absorption resonances. The particle dimer is
studied first as a tutorial example. For extended systems,
long-range $1/r$ dynamical interaction between polarized atoms or
particles is shown to give rise to soft modes observed in particular
in one-dimensional (1D) periodic arrangements. Finally, periodic
arrays confined within a 1D well are shown to exhibit imaginary
collective-motion eigenfrequencies for some ranges of the particles
spacing, thus revealing instabilities that preclude structural
transformations.

\section{Theoretical background}

We start by analyzing collective motion of small particles such as
atoms, molecules or nanoparticles that respond to electromagnetic
fields basically as induced dipoles driven by their polarizability
$\alpha(\omega)$, where $\omega$ is the light frequency. The
time-averaged force acting on one of such particles in the presence
of an external electric field $\Eb^{\rm ext}(\rb,t)=2\;\Re\{\Eb^{\rm
ext}(\rb)\exp(-\ii\omega t)\}$ is readily obtained from the integral
of Maxwell's stress tensor on a small sphere surrounding the
particle. One finds \cite{GA1980}
\begin{eqnarray}
   \Fb=2\;\Re\left\{\alpha\sum_l E^{\rm ext}_l\left[\nabla E^{\rm ext}_l\right]^*\right\}
      =-\nabla V-2\;\Im\{\alpha\}\,\Im\left\{\sum_l E^{\rm ext}_l\left[\nabla E^{\rm ext}_l\right]^*\right\}, \label{ee1}
\end{eqnarray}
where $l$ labels Cartesian coordinates, $V(\rb)=-\Re\{\alpha\}
|\Eb^{\rm ext}(\rb)|^2$ acts as an effective potential (i.e., the
particle can be trapped in regions of low or high electric field
strength when $\Re\{\alpha\}$ is negative or positive,
respectively), the second term in the right hand side of Eq.\
(\ref{ee1}) describes the so-called radiation pressure, and we use
Gaussian units throughout this paper.

Focusing on atoms, the frequency-dependent atomic polarizability
near a resonance at frequency $\omega_0$ with no decay channels
other than radiative takes a simple form compatible with the optical
theorem condition $\Im\{-1/\alpha\}=2k^3/3$ \cite{L1983}:
\begin{eqnarray}
   \alpha(\omega)=\frac{3c^3\Gamma/2\omega_0^2}
                  {\omega_0^2-\omega^2-\ii\Gamma\omega^3/\omega_0^2}
   \approx
                  \frac{3c^3\Gamma/4\omega_0^3}{\omega_0-\omega-\ii\Gamma/2},
                  \label{ee2}
\end{eqnarray}
where $k=\omega/c$ is the momentum of light in free space, $\Gamma$
is the resonance frequency width, and the last approximation is
valid for $\Gamma\ll\omega_0$ and $\omega$ near the resonance.
Furthermore, our analysis assumes low-enough light intensities to
exclude saturation effects near resonance, which would require to go
beyond linear response approximation.

When several particles placed in vacuum are considered, their
induced dipoles $\pb_j$ express the response to the external field
plus the field scattered by the other particles, that is,
\begin{eqnarray}
   \pb_j=\alpha \,\left[\Eb^{\rm ext}(\rb_j)+\sum_{j'\neq j} G(\rb_j-\rb_{j'}) \pb_{j'}\right],
   \label{ee22}
\end{eqnarray}
where $\rb_j$ labels particle positions and an $\exp(-\ii\omega t)$
time dependence is understood. The $3\times 3$ matrix
\begin{eqnarray}
   G(\rb)=\left(k^2+\nabla\otimes\nabla\right)\,\frac{\ee^{\ii k r}}{r}
         =A\,-\,\frac{B}{r^2}\;\rb\otimes\rb, \label{eqG}
\end{eqnarray}
with coefficients
\begin{eqnarray}
   A&=&\frac{\ee^{\ii kr}}{r^3}\,\left[(kr)^2+\ii kr-1\right],
   \label{AAA}\\
   B&=&\frac{\ee^{\ii kr}}{r^3}\,\left[(kr)^2+3\ii kr-3\right],
   \label{BBB}
\end{eqnarray}
describes the electric field produced at the position $\rb$ by a
dipole at the origin. All particles are considered to be equal for
simplicity. The force acting on the particle at $\rb_j$ is then
obtained from Eq.\ (\ref{ee1}) by adding the scattered electric
field $\sum_{{j'}\neq j} G(\rb-\rb_{j'}) \pb_{j'}$ to $\Eb^{\rm
ext}(\rb)$. Up to here, our analysis follows previous developments
of optical forces acting on small particles described through their
dipolar response \cite{ZKS04,G06}. This type of formalism can only
be applied to particles that are small with respect to both the
wavelength and their separation. However, its extension to include
multipolar terms in the interaction between neighboring particles
has been successfully carried out both for axially-symmetric
particles using 3D multipolar expansions \cite{paper089} and for
objects with translational invariance using 2D multipoles
\cite{GKK06_2,GKK06}.

It is easy to see that the dynamics can no longer be obtained from
an effective potential $V$, even in the absence of radiation
pressure. This is a manifestation of the fact that we are dealing
with open systems in which the photon bath provided by the external
field can add (remove) energy to (from) the motion of the particles.

\section{Optical dimer}

We shall discuss first a simple atomic dimer system that illustrates
the effect of strong inter-atomic interaction near an absorption
resonance. Optical biding in a dimer has been extensively studied
both theoretically \cite{DV94} and experimentally
\cite{BFG1989,MDW06}, but here we shall focus on the dynamical
aspects of such system.

\begin{figure}
\includegraphics[width=130mm,angle=0,clip]{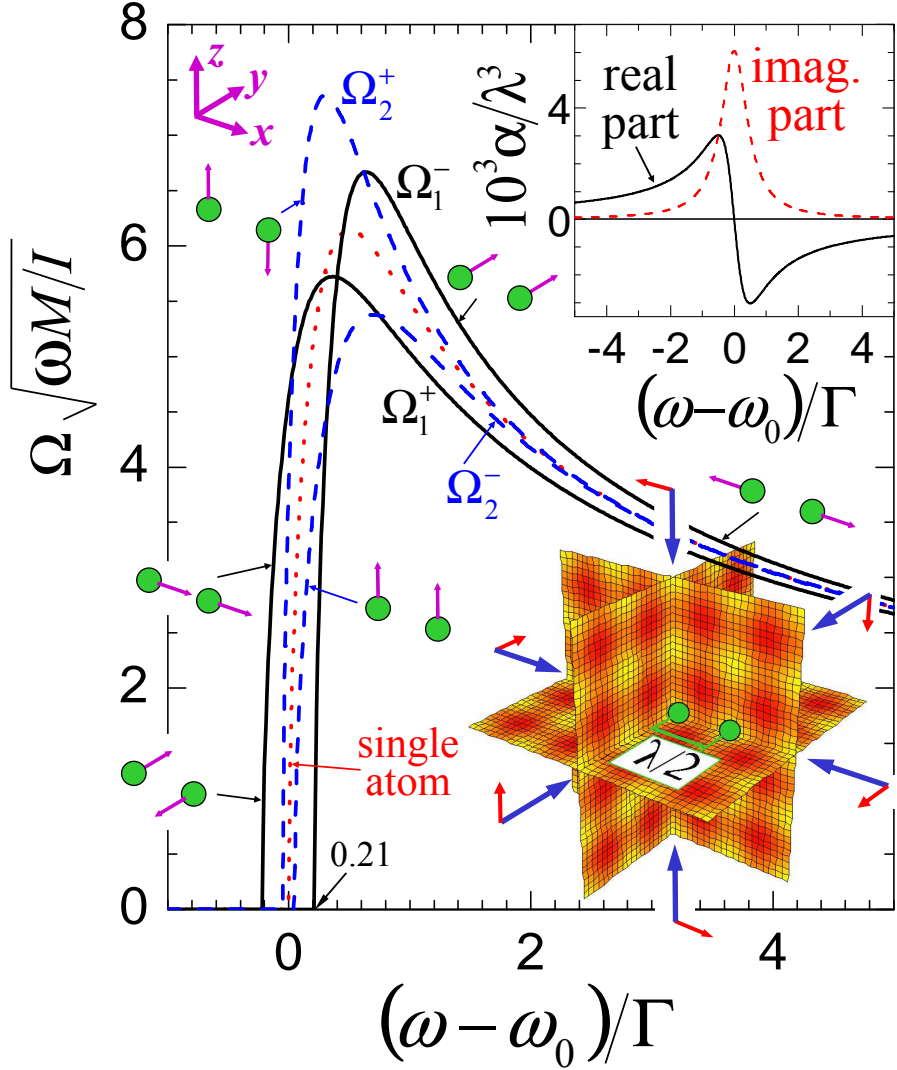}
\caption{Oscillation frequencies of two identical atoms trapped in
contiguous wells of an optical lattice as a function of light
frequency $\omega$ near an atomic resonance $\omega_0$. The lattice
is set up by three pairs of counter-propagating lasers of direction
and (linear) polarization as shown in the lower inset. The wells are
a distance $\lambda/2$ apart, where $\lambda=2\pi c/\omega$ is the
light wavelength. The atomic polarizability $\alpha$ (upper inset)
is obtained by assuming that radiative decay is the dominant
contribution to the resonance width $\Gamma$ [see Eq.\ (\ref{ee2})].
The oscillation frequency $\Omega$ is normalized using the atom mass
$M$ and the intensity of each beam $I$. The oscillations take place
around equilibrium positions corresponding to the vanishing of the
electric field strength (yellow regions in lower inset). Six
oscillation modes are obtained: two pairs of doubly-degenerate modes
(solid curves) and two nondegenerate modes (dashed curves). The
relative directions of motion of the atoms in the dimer are
indicated by arrows for each of the modes. The oscillation frequency
of a single trapped atom is given for reference (dotted curve).}
\label{Fig1}
\end{figure}

An optical lattice will be considered to be formed by three pairs of
equal-intensity counter-propagating lasers that define an orthogonal
$xyz$ frame, with their linear polarizations as shown in the lower
inset of Fig.\ \ref{Fig1}. The external electric field produced by
the lasers is then given by
\begin{eqnarray}
\Eb^{\rm ext}(\rb,t)=4\;\Re\left\{E_0\;\ee^{-\ii\omega\,t}\right\}
\;\left[\sin(kz)\;\xx+\sin(kx)\;\yy+\sin(ky)\;\zz\right], \nonumber
\end{eqnarray}
where $E_0$ is the electric field amplitude of each laser. For
$\omega>\omega_0$ [see Eq.\ (\ref{ee2})], the atomic polarizability
is essentially negative, as shown in the upper inset of  Fig.\
\ref{Fig1}, which represents the polarizability of Eq.\ (\ref{ee2})
scaled to the wavelength ($\alpha/\lambda^3$), so that the two atoms
under consideration can be trapped at contiguous minima of the
electric field (light regions in the lower inset).

The oscillation frequencies, as derived from Newton's equation for
infinitesimal displacements from equilibrium positions, reduce after
lengthy, straightforward algebra from Eqs.\ (\ref{ee1}) and
(\ref{ee22}) to
\begin{eqnarray}
\Omega_1^\pm=\frac{|E_0|k}{\sqrt{M}}\;\;\sqrt{8\;\Re\left\{\frac{-1}{\alpha^{-1}\pm A}\right\}} \label{Ome1}
\end{eqnarray}
and
\begin{eqnarray}
\Omega_2^\pm=\frac{|E_0|k}{\sqrt{M}}\;\;\sqrt{8\;\Re\left\{\frac{-1}{\alpha^{-1}\pm(A-B)}\right\}}, \label{Ome2}
\end{eqnarray}
where $M$ is the atomic mass, $A=-8(\pi^2+\ii\pi-1)/\lambda^3$, and
$A-B=16(\ii\pi-1)/\lambda^3$, according to the definitions of Eqs.\
(\ref{AAA}) and (\ref{BBB}) for a dimer spacing $r=\lambda/2$. The
light-frequency dependence of these oscillations is shown in Fig.\
\ref{Fig1}, along with schematic representations of directions of
motion for the modes, which illustrate how different orthogonal
directions are decoupled. Modes of frequency $\Omega_1^\pm$ and
$\Omega_2^\pm$ and doubly and singly degenerate, respectively, and
they are represented by solid and dashed curves.

The dimer oscillation frequencies deviate considerably from the
single atom case (dotted curve) as a result of inter-atomic
interaction driven by the $\omega_0$ resonance. Interestingly, this
interaction pushes the range of instability of the dimer (signaled
by some imaginary eigenfrequency $\Omega$) slightly towards the
$\omega>\omega_0$ region as compared to the single atom.

Optical forces scale linearly with light intensity, and therefore,
the oscillation frequencies are proportional to the strength of the
applied electric field. This allows us to represent those
frequencies normalized to the light intensity flux per laser,
$I=|E_0|^2c/2\pi$. For instance, single Rb atoms ($M=85.46$ amu)
near their 780 nm resonance ($\Gamma=2\pi\times 4$ MHz
\cite{SWC03,note100}) will oscillate with frequency $\Omega\approx
0.36$ MHz for $I=10$ mW/cm$^2$ lasers tuned to
$\omega=\omega_0+2\Gamma$, which corresponds to a temperature of
$\approx 17$ $\mu$K. Notice that $\Omega\ll\Gamma$, so that Doppler
shifts can be safely neglected at this relatively high laser
intensity, for which the atomic motion can be described classically,
although saturation effects could be an issue \cite{A1980}.

\section{Periodic arrays}
\label{periodicarrays}

Oscillations in periodic configurations of extended arrays can be
studied by introducing small perturbations in Eq.\ (\ref{ee22}) for
specific momentum $\qb$, corresponding to a displacement of every
particle $j$ around its equilibrium position $\rb_j$ given by
$\ub_j=\ub_\qb\exp(\ii\qb\cdot\rb_j)$. For small perturbations, the
electromagnetic forces scale linearly with the displacement
$\ub_\qb$, so that one obtains an equation of motion of the form
\begin{eqnarray}
\Re\{\Sigma_\qb\}\ub_\qb=-M\Omega^2_\qb\ub_\qb, \label{Newton}
\end{eqnarray}
where $M$ is the particle mass. Three branches of the oscillation
frequency $\Omega_\qb$ are obtained in general out of the
eigenvalues of the force constants matrix $\Sigma_\qb$.

\subsection{1D arrays}

We shall first focus on blue-detuned systems, in which the particle
polarizability has negative real part, so that the unperturbed
particles sit at light intensity minima. Then, the force constants
matrix reduces to
   \begin{eqnarray}
      \Sigma_\qb = |E_0|^2k^2\;\;{\bf C}^+ \frac{1}{\frac{1}{\alpha}-G_\qb} {\bf C},
   \nonumber
   \end{eqnarray}
where
   \begin{eqnarray}
      G_\qb = \sum_{j\neq 0} \ee^{-\ii\qb\cdot\rb_j} G(\rb_j) \label{ee3}
   \end{eqnarray}
is the sum of the dipole-dipole interaction extended over lattice
sites $\rb_j$, and ${\bf C}$ is a dimensionless matrix that depends
on the specific orientations and polarizations of the lasers setting
up the optical lattice.

In particular, we shall consider a 1D infinite string formed by the
same kind of blue-detuned atoms as in Fig.\ \ref{Fig1}, with one
atom trapped at each minimum of the electric-field along the
$\langle 111\rangle$ direction of the same optical lattice. The
oscillation frequencies are then labeled by the longitudinal
momentum $q$. Displacements parallel ($\parallel$) and perpendicular
($\perp$) to the array turn out to be decoupled. We find the
following analytical expression for the oscillation frequencies, similar in structure
to Eqs.\ (\ref{Ome1}) and (\ref{Ome2}):
\begin{eqnarray}
\Omega_q^\sigma=\frac{|E_0|k}{\sqrt{M}}\;\;\sqrt{8\;\Re\left\{\frac{-1}{\alpha^{-1}-G_q^\sigma}\right\}},
\nonumber
\end{eqnarray}
where $\sigma=\parallel,\perp$ in each case, the lattice sums reduce
to
\begin{eqnarray}
   G_q^\parallel=4\sum_{j=1}^\infty\cos(qaj)\frac{\ee^{\ii kaj}}{(aj)^3}(1-\ii kaj)
   \label{fqpara}
\end{eqnarray}
and
\begin{eqnarray}
   G_q^\perp=2\sum_{j=1}^\infty\cos(qaj)\frac{\ee^{\ii kaj}}{(aj)^3}[(kaj)^2+\ii
   kaj-1],
   \label{fqperp}
\end{eqnarray}
and $a=\sqrt{3}\lambda/2$ is the period of the array, so that
$ka=\sqrt{3}\pi$.

\begin{figure}
\includegraphics[width=130mm,angle=0,clip]{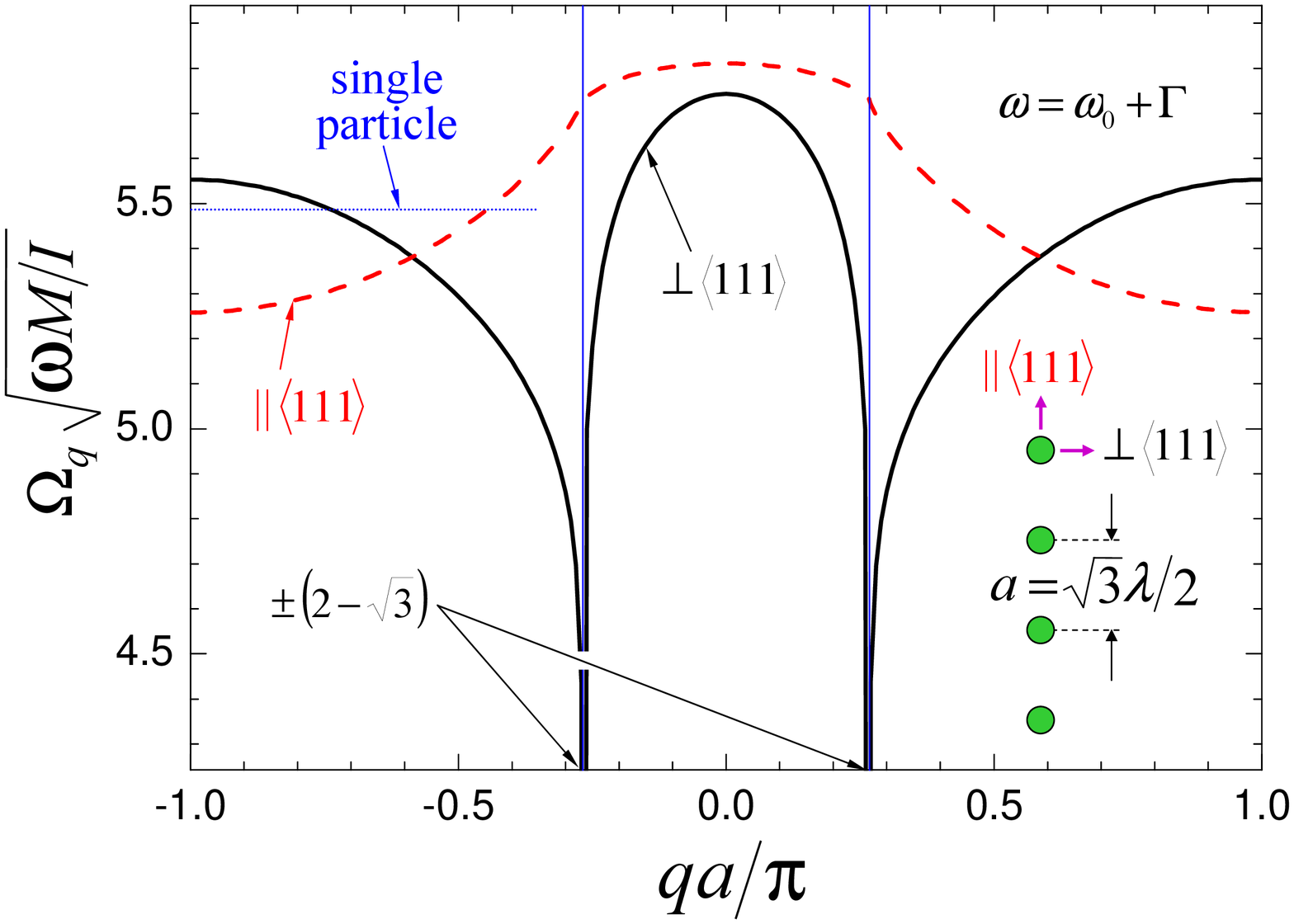}
\caption{Dispersion relation of the oscillation modes in an infinite
periodic string of atoms located at the minima of electric field
strength along the $\langle 111\rangle$ direction under the same
conditions of illumination as in Fig.\ \ref{Fig1} (see lower inset
in that Fig.; the coordinate axes are taken parallel to the laser
beams). The nearest-neighbor distance is $a=\sqrt{3}\lambda/2$. The
light frequency is tuned to $\omega=\omega_0+\Gamma$, using the same
notation and atomic polarizability as in Fig.\ \ref{Fig1}. The
oscillation frequency is given as a function of momentum $q$ along
the string within the one-dimensional first Brillouin zone.
Oscillations parallel to the $\langle 111\rangle$ direction (broken
curve) are decoupled from those along perpendicular directions
(solid curve).} \label{Fig2}
\end{figure}

Fig. \ \ref{Fig2} shows results obtained using this formalism within
the first Brillouin zone in $q$ space. Soft modes of zero frequency
can be observed in the degenerate oscillations along directions
transversal to the string, giving rise to large group velocities.
This results from phase accumulation in the long-range inter-atomic
interaction within Eq.\ (\ref{fqperp}), which diverges as
$\ln|\varphi\pm q a|$ near $\mp q a=\varphi=(2-\sqrt{3})\pi$. This
logarithmic divergence is obviously destroyed when the string is
finite, regardless its length, and therefore, it cannot lead to
permanent distortions of the lattice from the equidistant array
configuration.

\subsection{Beyond the dipole approximation in 2D arrays}

For particles of non-negligible size compared to the wavelength,
multipoles beyond dipoles may become important. This situation can
be described by a relation similar to Eq.\ (\ref{ee22}), where $\pb$
is then understood as a vector of multipolar amplitudes and $\alpha$
is the multipolar scattering matrix \cite{paper040}. Analyses along
these lines have been previously offered by the author for momentum
transfer from a fast electron to a particle \cite{paper093} and for
neighboring effects in the interaction force between two
arbitrarily-shaped particles \cite{paper089}. Higher-order
multipoles beyond the dipole yield a more complex expression for the
force, but the quadratic dependence on the field exhibited by Eq.\
(\ref{ee1}) is still maintained.

\begin{figure}
\includegraphics[width=130mm,angle=0,clip]{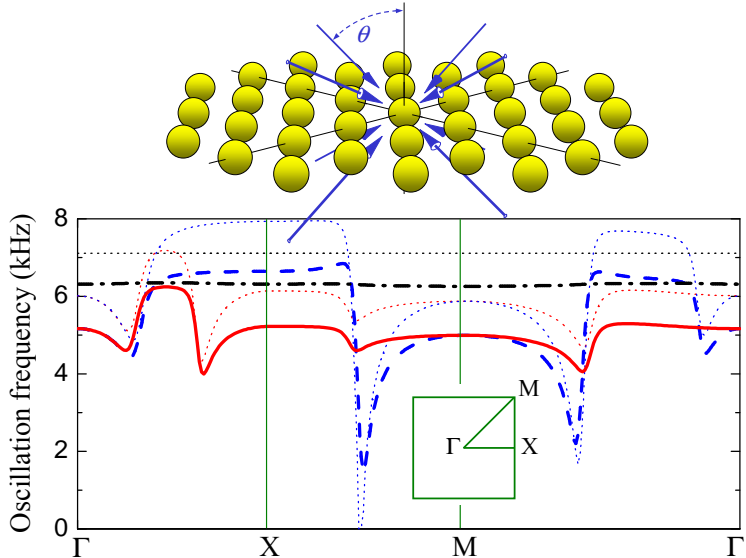}
\caption{Dispersion relation of oscillation modes in an optical
lattice of silica spherical particles ($\epsilon=2.1$) in vacuum.
The lattice is set up by four pairs of counter-propagating
linearly-polarized beams with incident magnetic field in the plane
of the particles. The inset shows the particles in their equilibrium
positions forming a planar square lattice of nearest-neighbor
distance $\lambda/\sin\theta$, where $\theta=50^\circ$ is the angle
formed by the light beams and the normal to the plane of the array.
Continuous, dashed, and dashed-dotted thick curves correspond to the
three oscillation modes for each value of the parallel momentum
along the excursion $\Gamma$XM$\Gamma$ within the first Brillouin
zone. The spheres diameter is 240 nm, the light wavelength is
$\lambda=550$ nm, and the flux of each laser is 100 W/cm$^2$. These
results are obtained with inclusion of higher-order multipoles
beyond the dipole. Calculations in the dipole approximation (thin
dotted curves) are shown for reference, with the polarizability obtained
from the electric dipole Mie coefficient as $\alpha=(3/2k^3)\;t_1^E$ \cite{paper040}.} \label{Fig3}
\end{figure}

These conditions have been considered in Fig.\ \ref{Fig3} for a
two-dimensional (2D) square array of silica beads (see upper inset)
trapped in a field given by
\begin{eqnarray}
\Eb^{\rm ext}(\rb,t)=8\;\Re\left\{E_0\;\ee^{-\ii\omega\,t}\right\}
\;\Big\{\left[\sin(Qx)\;\xx+\sin(Qy)\;\yy\right]\sin(k_zz)\;\cos\theta
\nonumber\\
\;+\;\left[\cos(Qx)+\cos(Qy)\right]\;\zz\;\cos(k_zz)\;\sin\theta\Big\},
\nonumber
\end{eqnarray}
where $\theta$ is the angle between the beam directions and the
normal to the plane of the array, $Q=k\sin\theta$, and
$k_z=k\cos\theta$. This system is experimentally challenging because
it requires to trap particles in vacuum, for which pioneering
results were reported three decades ago by Ashkin and Dziedzic
\cite{AD1976}, who achieved optical levitation in high vacuum.

Three oscillation frequency bands are obtained and represented for
an excursion along symmetry points within the first Brillouin zone
of the reciprocal lattice in momentum space. One of the bands
corresponds to nearly-independent-particle motion perpendicular to
the plane of the array and is almost dispersionless (dashed-dotted
thick curve). The remaining two bands (solid and dashed thick
curves) are degenerate at the $\Gamma$ and M points, which is a
condition imposed by symmetry.
% Light-induced forces can be enhanced near plasmons in metallic nanoparticles \cite{HBH05,PLK06}, which allow mimicking the situation encountered near atomic resonances.
Unlike the previous examples, the dielectric particles of Fig.\
\ref{Fig3} are attracted towards maxima of the light intensity (this
is the equivalent of red-detuned atomic resonances). The effect of
multipolar interaction is quantitatively significant (cf.
oscillation modes calculated in the dipole approximation,
represented by thin dotted curves in Fig.\ \ref{Fig3}; incidentally,
these latter dipole calculations should describe well red-detuned
atoms trapped in this optical-lattice geometry).

Low-frequency modes arising from long-range interactions are visible
in Fig.\ \ref{Fig3} as pronounced dips that become very sensitive to
finite array boundaries and size distribution of the particles.
These effects have been phenomenologically described in Fig.\
\ref{Fig3} through a small attenuation in the interaction between
distant particles represented by a dielectric function equal to
$1+0.03\ii$ in the surrounding medium, which gives an interaction
decay-length of $\sim 8$ times the nearest-neighbor distance.

\section{Phase transitions in 1D optical matter}

\begin{figure}
\includegraphics[width=120mm,angle=0,clip]{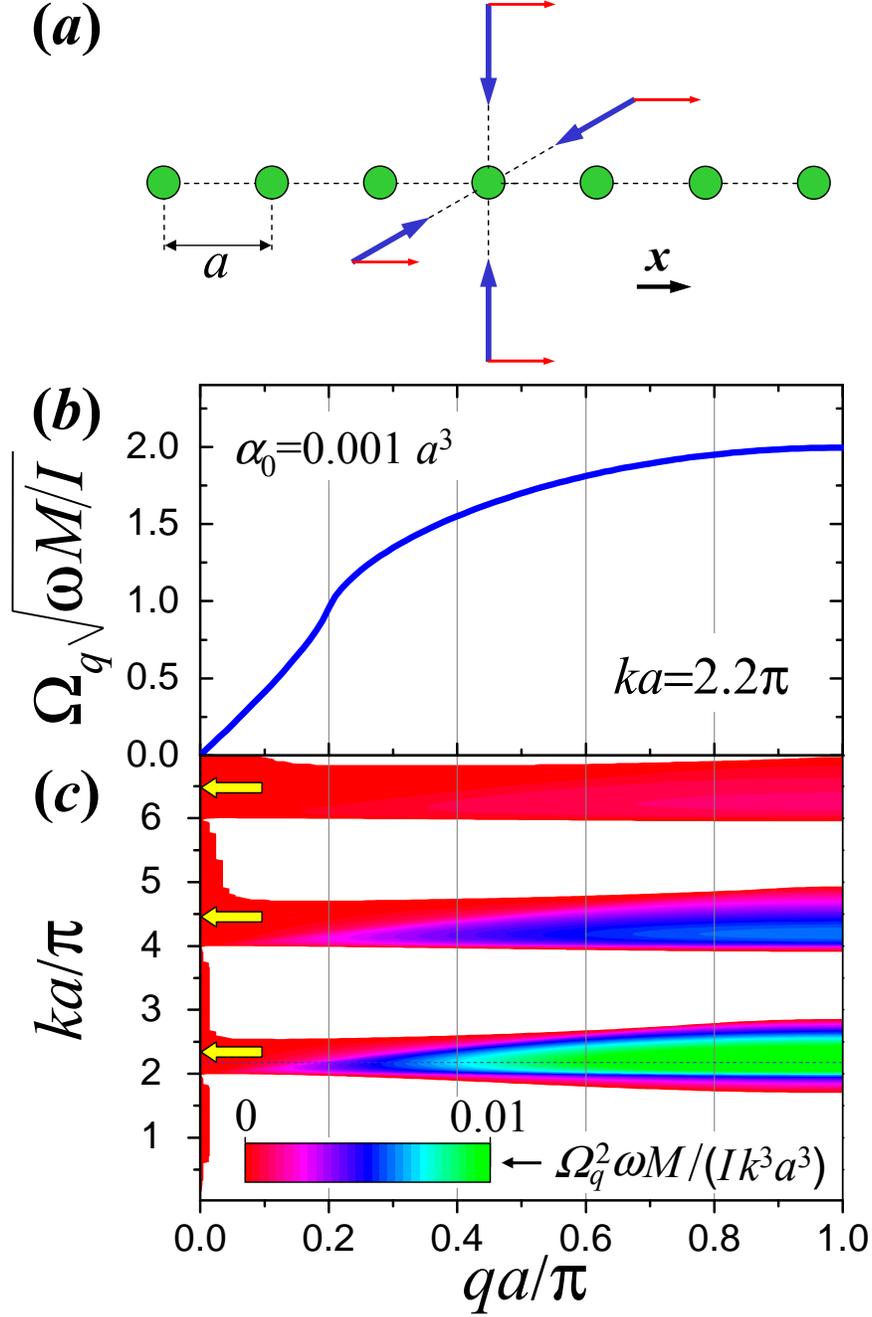}
\caption{Oscillation modes and stability of a linear array of
particles trapped in a two-dimensional optical lattice set up by two
pairs of counter-propagating beams with polarizations as shown in
(a). The array is assumed to be periodic, infinitely long, and
surrounded by vacuum. A typical spectrum of oscillations is
presented in panel (b) for specific values of the free-space
momentum of the trapping light $k$ and the particle polarizability
$\alpha$, normalized to the lattice spacing $a$ (see text insets).
The spectrum is given within the first Brillouin zone of momentum
along the array $q$, with normalization of frequency and momentum as
in Fig.\ \ref{Fig2}. The contour plot of panel (c) represents
squared oscillation frequencies as a function of $q$ and $k$ for
$\alpha=1/(\alpha_0^{-1}-2\ii k^3/3)$, where $\alpha_0=0.001 a^3$ is
the electrostatic polarizability (this prescription for $\alpha$ is
consistent with the optical theorem for non-absorbing particles
\cite{DF94}). White regions in (c) correspond to imaginary
oscillation eigenfrequencies, which signal array instabilities.}
\label{Fig4}
\end{figure}

The above analysis has been limited to particles trapped at the
sites of 3D lattices and capable of oscillating around their
equilibrium positions. We shall now explore a different system
comprised by particles that are confined along two spatial
directions but free to move along the remaining third direction.
This scenario is presented for instance in particles trapped inside
a cylindrical optical cavity (e.g., a hole of a photonic crystal
fiber), and also in particles confined to a 1D well of a 2D optical
lattice like that of Fig.\ \ref{Fig4}(a), in which the external
field is
\begin{eqnarray}
   \Eb^{\rm ext}(\rb,t)=4\;\Re\left\{E_0\;\ee^{-\ii\omega\,t}\right\} \;\left[\cos(kz)+\cos(ky)\right]\;\xx, \nonumber
\end{eqnarray}
where $E_0$ is the field amplitude of each laser. The particles will
be assumed to have positive polarizability $\alpha$ (i.e., the
confinement occurs in regions of maximum electric field) and to be
periodically spaced. This is a configuration of equilibrium for an
infinite chain, in which the force acting on the particles vanishes.
We shall then analyze the spectrum of collective oscillations within
the first Brillouin zone of the reciprocal lattice and explore the
stability of that equilibrium.

The self-consistent electric field amplitude acting on a given
particle reduces from Eq.\ (\ref{ee22}) to the analytical expression
\begin{eqnarray}
   E=\frac{4E_0}{1-\alpha G_{q=0}^\parallel}, \nonumber
\end{eqnarray}
where $G_{q=0}^\parallel$ [see Eq.\ (\ref{fqpara})] is the sum of
the fields scattered by the rest of the particles, and only the
$\exp(-\ii\omega t)$ component is considered. Small displacements
out of equilibrium along small vectors ${\bf u}_j$, as defined at
the beginning of Sec.\ \ref{periodicarrays}, will then modify the
self-consistent field in the vicinity of every particle $j$ to
\begin{eqnarray}
   \Eb_j=\Eb^{\rm ext}+{\bf \eta}_j, \nonumber
\end{eqnarray}
from which we discount the field scattered by that particle. The
field perturbations can be conveniently obtained by expanding
$G(\rb)$ [see Eq.\ (\ref{eqG})] in Taylor series and by working in
Fourier space $q$ according to the definition
\begin{eqnarray}
   {\bf \eta}_q=\sum_j\ee^{-\ii qaj}{\bf \eta}_j,\nonumber
\end{eqnarray}
similar to Eq.\ (\ref{ee3}). We find
\begin{eqnarray}
   \eta^\parallel_q=\frac{g_qE}{\frac{1}{\alpha}-G_q^\parallel}\;\;u_q^\parallel
   \nonumber
\end{eqnarray}
and
\begin{eqnarray}
   {\bf \eta}^\perp_q=\frac{(-g_qE/2)}{\frac{1}{\alpha}-G_q^\perp}\;{\bf u}^\perp_q,
   \nonumber
\end{eqnarray}
where $\parallel$ and $\perp$ indicate directions parallel ($\xx$)
and perpendicular ($\yy$, $\zz$) to the array, respectively,
$G_q^\parallel$ and $G_q^\perp$ were defined in Eqs.\ (\ref{fqpara})
and (\ref{fqperp}), and
\begin{eqnarray}
   g_q=4\ii\sum_{j=1}^\infty\sin(qaj)\frac{\ee^{\ii kaj}}{(aj)^4}[(kaj)^2+3\ii kaj-3]. \nonumber
\end{eqnarray}
Finally, the oscillation frequencies $\Omega_q$ are obtained from
Newton's equation, similar to Eq.\ (\ref{Newton}), but with a more
complex expression for $\Sigma_\qb$. Like in the system of Fig.\
\ref{Fig2}, oscillations along parallel and perpendicular directions
with respect to the array are decoupled and their frequencies reduce
to
\begin{eqnarray}
   \Omega_q^\parallel=\left[\frac{2|\alpha E|^2}{M}\;\;
                      \Re\left\{\frac{g_q^2}{1/\alpha-
                      G_q^\parallel}+H_q^\parallel\right\}\right]^{1/2} \label{aaaa}
\end{eqnarray}
and
\begin{eqnarray}
   \Omega_q^\perp=\left[\frac{2|\alpha E|^2}{M}\;\;
                      \Re\left\{\frac{g_q^2/4}{1/\alpha-
                      G_q^\perp}+H_q^\perp\right\}
                      +\frac{4k^2}{M}\Re\{E_0^*\alpha E\}\right]^{1/2}, \label{bbbb}
\end{eqnarray}
respectively, where
\begin{eqnarray}
   H_q^\parallel=4\sum_{j=1}^\infty\left[\cos(qaj)-1\right]\frac{\ee^{\ii kaj}}{(aj)^5}[\ii(kaj)^3-5(kaj)^2-12\ii kaj+12] \nonumber
\end{eqnarray}
and
\begin{eqnarray}
   H_q^\perp=8\sum_{j=1}^\infty\left[\cos(qaj)-1\right]\frac{\ee^{\ii kaj}}{(aj)^5}[(kaj)^2+3\ii kaj-3]. \nonumber
\end{eqnarray}
Perpendicular modes span a nearly-flat band around a high central
frequency determined by the confining transversal potential [see the
right-most term inside the square brackets of Eq.\ (\ref{bbbb})].
The stability of the array will be exclusively determined by the
band of longitudinal modes. A typical dispersion relation is shown
in Fig.\ \ref{Fig4}(b) for specific values of the polarizability
(see inset) and spacing-to-wavelength ratio ($a/\lambda=1.1$). In
contrast to the system of Fig.\ \ref{Fig2}, the longitudinal band is
acoustic, so that the oscillation frequency vanishes in the
$q\rightarrow 0$ limit (notice that $g_q=H_q=0$ in this limit),
standing for a rigid slow translation of all particles along the 1D
trapping well.

The stability of this type of array is analyzed in Fig.\
\ref{Fig4}(c), which represents the oscillation spectrum given by
Eq.\ (\ref{aaaa}) as a function of lattice spacing. White regions
correspond to modes of imaginary frequency that describe motion
beyond the validity of the harmonic approximation. The presence of
these modes signals structural instabilities for some ranges of the
period of the array, which must rearrange itself by creating defects
with different particle separation. However, there are certain
ranges of spacings that are stable, centered around the distance of
equilibrium for a particle dimer in the channel (see horizontal
arrows). This analysis of the array stability is also valid in the
presence of friction forces, like those that would show up if the
particles were immersed in a fluid. A practical realization of such
arrays using holographic tweezers \cite{G03,GR06} combined with
transversal confining light would allow one to study their stability
and to assess mechanical properties like the compressibility by
fixing the position of the particles at the end of a finite array.

\section{Conclusion}

Atoms and nanoparticles trapped in optical lattices have been shown
to exhibit collective oscillations, the spectra of which reveal
complex patterns that can be controlled by external illumination
conditions. The oscillation frequency increases with the amplitude
of the binding lasers, thus adding an extra degree of freedom that
allows obtaining frequencies in the kHz-MHz range in the case of
atomic lattices. These oscillations constitute a genuine form of
dynamical collective behavior in optical matter encompassing complex
phenomena such as soft-modes and lattice phase transitions.
Furthermore, our analysis of the oscillation spectra has been shown
to resolve the question of the stability of the arrays.

The results presented here could be useful to design the following
specific experiments:
\begin{itemize}
\item Collective motion in optical lattices of trapped atoms,
which should be observable in their response to low-frequency
radiation within the noted oscillation frequency range.

\item Stability of 1D nanoparticle chains trapped along one channel of a 2D
optical lattice. Initial particle trapping is possible using
holographic illumination \cite{GR06}, superimposed to the optical
lattice, and the stability could be proved by varying the relative
intensity of holographic and optical-lattice laser beams.

\item Phase transitions in optical matter, using a similar setup as
in the previous point, but holding the ends of a chain at specific
locations. Transitions should occur as these ends are displaced.
\end{itemize}

Finally, there are a number of open questions posed by these novel
oscillations that will require separate developments, including the
extension of this work to lower laser intensities and to conditions
for which the oscillations behave as quasi-particles in atomic
lattices. Interesting phenomena could also result from saturation
effects at large laser intensities, whereby the atomic response is
far from the linear regime. Equally important, thermodynamics of
open systems like the arrays of Fig.\ \ref{Fig4} is expected to
yield new properties associated to physical quantities such as the
compressibility when exerting pressure on finite strings or the
specific heat associated to the oscillations. These are challenges
that will ultimately determine the applicability of collective
behavior of optical matter to fields such as quantum information
technology and signal processing.

\section*{Acknowledgments}

The author wants to thank M. Nieto-Vesperinas for enjoyable
discussions. This work has been supported in part by the Spanish MEC
(contracts FIS2004-06490-C03-02 and NAN2004-08843-C05-05) and by the
EU (STREP STRP-016881-SPANS and NoE Metamorphose).

%\bibliographystyle{apsrev}
%\bibliography{../../../bibtex/refs}

\begin{thebibliography}{33}
\bibitem[{not()}]{note100}
\bibinfo{note}{It should be noted that Rb has another resonance of width
  $\Gamma=2\pi\times 2$ MHz at 795 nm (frequency $\omega_1$), which combined
  with the 780 nm resonance (frequency $\omega_0$) gives an effective value of
  $\Gamma=2\pi\times 6$ MHz for light tuned far from this region
  ($|\omega-\omega_0|\gg\omega_0-\omega_1$). However, we are discussing here
  light tuned very close to the 780 nm resonance
  ($|\omega-\omega_0|\ll\omega_0-\omega_1$), for which the lower-frequence
  resonance can be overlooked.}
\end{thebibliography}

\end{document}